\documentclass[times,twocolumn,final,authoryear]{elsarticle}

\usepackage{jasr}
\usepackage{framed,multirow}
\usepackage{amssymb}
\usepackage{latexsym}

\usepackage{url}
\usepackage{xcolor}
\definecolor{newcolor}{rgb}{.8,.349,.1}

\usepackage[pdfstartview=FitH, hidelinks]{hyperref}

\usepackage{amsmath}
\newcommand{\dd}{\mathrm{d}}

\newcommand{\rev}[1]{#1}

\journal{Advances in Space Research}

\begin{document}

\verso{Olivier Herscovici-Schiller \textit{et al.}}

\begin{frontmatter}

\title{A simple ionospheric correction method for radar-based space surveillance systems, with performance assessment on GRAVES data}

\author[1]{Olivier \snm{Herscovici-Schiller}\corref{cor1}}
\cortext[cor1]{Corresponding author}
  \ead{olivier.herscovici@onera.fr}
\author[1]{Fabien \snm{Gachet}}
\author[1]{Jocelyn \snm{Couetdic}}
\author[1]{Luc \snm{Meyer}}
\author[1]{S\'ebastien \snm{Reynaud}}

\address[1]{DTIS, ONERA, Universit\'e Paris Saclay, F-91123 Palaiseau, France}

\received{}
\finalform{}
\accepted{}
\availableonline{}
\communicated{}

\begin{abstract}
  Ionospheric effects degrade the quality of radar data, which are critical for the precision of the satellite ephemeris produced by space surveillance systems; this degradation is especially noticeable for radars such as GRAVES that operate in the very high frequency range.
  This article presents a simple and effective method to correct for ionospheric effects, with an evaluation on data obtained with GRAVES, the French space surveillance radar.
  This method relies on GPS data, and our evaluation relies on GRAVES and DORIS data.
  We found that the gain in terms of evaluated radial velocity can be as high as 1.76$\kappa$, where $\kappa$ is the typical root mean square of the noise on radial velocity measurements for GRAVES (excluding ionospheric effects): the error decreases from 2.60$\kappa$ to 0.83$\kappa$ for daytime satellite overhead passes.
  Our conclusion is that, while this method is very simple to implement, it has proven to be a good correction for ionospheric effects in practice.
\end{abstract}


\end{frontmatter}

\section{Introduction}
  Radar-based space surveillance systems are an important part of any space situational awareness programme.
  The quality of the data is critical for the precision of the satellite ephemeris produced by the system.
  An important source of error in the data is the influence of the ionosphere.

  Engineers who design a space surveillance radar thus have to measure and compensate the perturbation on the radar wave that is induced by its propagation in the ionosphere~\rev{\citep{hapgood_ionospheric_2010}}.

  The aim of this article is to describe a robust and very easy-to-use method to take into account the influence of the ionosphere, along with an assessment of the performance of this method using data acquired by GRAVES (grand r\'eseau adapt\'e \`a la veille spatiale) --- the French space surveillance radar --- on DORIS (Doppler Orbitography by Radiopositioning Integrated on Satellite) satellites. This method relies only on two dual-frequency GNSS (Global Navigation Satellite System) receivers: one at the emission site of the radar, and one at the reception site. \rev{Our goal is to allow an estimation of the total electronic content of the ionosphere using only a single station at the reception site and a single station at the reception site.}

  In section~\ref{sec:ionosphere}, we describe the influence of the ionosphere on electromagnetic waves.
  In section~\ref{sec:gnsstotec}, we introduce how we retrieve the total electron content from GNSS data.
  In section~\ref{sec:doris}, we assess the gain in accuracy on the radial velocity. We do this by taking DORIS data as the ground truth, and use it to calculate the error that is made in the evaluation of the radial velocity measurement with and without the correction.

\section{The impact of the ionosphere on the data collected by a space surveillance radar}
\label{sec:ionosphere}
  \subsection{The ionosphere}
    The ionosphere is the part of the atmosphere that is ionized, mainly under the influence of the Sun. A good description of the ionosphere from the point of view of ionospheric tomography is given by~\cite{fehmers_tomography_1996}. The quantity of interest is the total electron content (TEC). It is the linear integral of the electron density in a given direction. The total electron content from the ground to the zenith at any point of the Earth is called the vertical total electron content (vTEC). The total electron content in a given direction, for example from a ground station to a satellite, is called the slant total electron content (sTEC). The TEC (or vTEC, or sTEC) is measured in~m$^{-2}$ in the international system, but it is often expressed in TECu (total electron content unit), with 1~TECu~$= 10^{16}$~m$^{-2}$.

    A review of the techniques for modeling the total electron content of the ionosphere can be found in~\cite{bust_history_2008}. \rev{New developments are still ongoing (see for example \cite{bidaine_assessment_2010,ansari_ionospheric_2017,li_high-precision_2020,yasyukevich_estimating_2015,yasyukevich_gnss-based_2020,yasyukevich_mitigator_2022,li_satellite-based_2021,pudlovskiy_ionospheric_2021}); notably machine learning techniques have developed in this field like many others~\citep{orus_perez_using_2019,mallika_i_machine_2020,ferreira_short-term_2017}. Typical examples of tomography reconstruction algorithm of the ionosphere in the literature, such as~\cite{seemala_three-dimensional_2014}, model the TEC over wide timescales (more than an hour, possibly more than a half day) and wide geographical regions (with cells hundreds of kilometers wide).} The goal of such methods is usually to model the ionosphere as precisely as possible, using as many data point as possible. This is quite different from our goal here, which is to model the impact of the ionosphere as seen from a single ground station as easily as possible, using only data from the station.

    Among the models of the ionosphere in the context of GNSS, two notable ones are the Klobuchar model, and NeQuick. The Klobuchar model~\citep{klobuchar_ionospheric_1987}, which is particularly simple~\citep{petit_iers_2010}, is used to correct the signal received from GPS (Global Positioning System) satellites. The NeQuick model~\citep{nava_new_2008}, is used to correct the signal received from Galileo satellites. \rev{There are many more models available. The interested reader can refer to~\cite{collective_guide_2014}.}

  \subsection{The propagation of electromagnetic waves in the ionosphere}
    The propagation of electromagnetic waves in the ionosphere is affected by the medium: electromagnetic waves are delayed and their phases are shifted. This induces a shift in the Doppler measurement $\Delta v$ of the radial velocity of the satellite as seen by the radar, which is well approximated~\citep{parkinson_global_1996} by

    \begin{equation}
      \Delta v = \frac{40.3}{f^2}\times \frac{\dd \rev{I_s}}{\dd t},
    \end{equation}
    where $f$ is the frequency of the radar wave, $\rev{I_s}$ is the sTEC (slant total electron content) and $t$ is the time.
    The factor $40.3$ is correct as long as all parameters are expressed in the units of the international system.
    A key point for ionospheric tomography is that this phase shift is chromatic, which allows the use of dual-band GNSS receivers to retrieve the electron content \citep{odijk_ionosphere-free_2003} of the ionosphere. \rev{Our theoretical framework for the retrieval of the slant electron content is close to the one used by~\cite{lanyi_comparison_1988}.}

\section{The retrieval of the electron content from GNSS data}
\label{sec:gnsstotec}
  \subsection{The forward model}
    Let us call $\Delta P$ the difference between the two pseudo-distance signals emitted by a GNSS satellite, measured from a ground receiver. According to a classic equation, well explained in \cite{kunitsyn_ionospheric_2003} (section 6.2),

    \begin{equation}
      40.3 \times \frac{f_1^2 - f_2^2}{f_1^2 \times f_2^2} \times \Delta P = \rev{I_s} + b_s + b_r,
    \end{equation}
    where $f_1$ and $f_2$ are the two frequencies of the two different signals, $b_s$ is the bias of the satellite, and $b_r$ is the bias of the receiver. \rev{Following~\citep{el-rabbany_introduction_2011}, we will regard those biases as constants over a few hours, although they may sometimes change faster~\citep{li_multi-frequency_2020}.}

    We chose to characterize the ionosphere by a single coefficient, which is the vTEC at the zenith over the ground station. We follow~\cite{klobuchar_ionospheric_1987} and choose a single-layer representation of the ionosphere. The relationship between the vTEC and the sTEC ($\rev{I_s}$) is then

    \begin{equation}
      \rev{I_s}(e) =  \dfrac{\rev{I_v}}{\rev{\cos}\left\{\sin^{-1}\left[\frac{R_T}{R_T + h} \times \cos(e) \right] \right\}},
    \end{equation}
    where $e$ is the elevation of the satellite, $\rev{I_v}$ is the vTEC, $R_T$ is the radius of the Earth and $h = 350$\,km is the altitude of the ionosphere layer. \rev{Using the fact that $\cos = \pm \sqrt{1 - \sin^2}$, this is equivalent to the equation~(3) of \cite{wielgosz_validation_2021} (for $\alpha = 1$), itself building upon \cite{schaer_mapping_1999}.}
    Finally, our forward model reads as follows: for each satellite $s$, for each time step $t$,

    \begin{align}
      40.3 \times & \frac{f_1^2 - f_2^2}{f_1^2 \times f_2^2} \times \Delta P(s,t) =  \nonumber \\
            & \dfrac{\rev{I_v}}{\rev{\cos}\left\{\sin^{-1}\left[\frac{R_T}{R_T + h} \times \cos(e) \right] \right\}} + b_s(s) + b_r .
            \label{eq:modeledirect}
    \end{align}

  \subsection{The inversion}
    Since the noise is white, and the dependence of the unknown parameters on the data has been constructed in order to be linear, we take our estimate $\widehat{\rev{I_v}}$ as the least-squares solution of the observation system.
    We write the forward model, equation~(\ref{eq:modeledirect}), as :

    for each satellite $s$, for each time step $t$,

    \begin{equation}
      \Delta P(s,t) = C_1(s,t)  \times \rev{I_v} + C_2 \times b_s(s) + C_2 \times b_r,
      \label{eq:modelelineaire}
    \end{equation}
   where the two known parameters $C_1(s,t)$ and $C_2$ are defined as

   \begin{align}
     C_1(s,t) & = \frac{f_1^2 \times f_2^2}{40.3\times(f_1^2 - f_2^2)} \nonumber \\
         & \quad \times  \dfrac{\rev{1}}{\rev{\cos}\left\{\sin^{-1}\left[\frac{R_T}{R_T + h} \times \cos(e) \right] \right\}} \\
     C_2 & = \frac{f_1^2 \times f_2^2}{40.3\times(f_1^2 - f_2^2)}.
   \end{align}
   This is a linear system of type

   \begin{equation}
     Y = AX.
   \end{equation}
   Let us call $S$ the total number of satellites in the observation timeframe, and $T$ the number of time steps in the observation timeframe.
   $Y$ is the observation vector. It contains the $S \times T$ measurements $\Delta P(s,t)$. $X$ is the unknown vector. Its transpose $X^\dagger$ is such that

   \begin{equation}
     X^\dagger = \left[ \widehat{\rev{I_v}} , b_s(s = 1), ... \; , b_s(s=S), b_r  \right].
   \end{equation}
   $A$ is the matrix of the system. Considering equation~(\ref{eq:modelelineaire}),

   \begin{equation}
     A =
     \begin{bmatrix}
       C_1(s = 1, t = 1) & C_2 & 0 & ... & 0 & C_2 \\
       C_1(s = 1, t = 2) & C_2 & 0 & ... & 0 & C_2 \\
       \vdots & \vdots & \vdots & ... & \vdots & C_2 \\
       C_1(s = S, t = T) & 0 & ... & 0 & C_2 & C_2
     \end{bmatrix}.
   \end{equation}
   A general line of matrix $A$, for satellite number $s = \sigma$, for timestep $t = \tau$, is given by

   \begin{equation}
     A_{\sigma, \tau} =
     \begin{bmatrix}
        C_1(s = \sigma, t = \tau) & 0 & ... & 0 & C_2 & 0, ..., 0, C_2
    \end{bmatrix}.
   \end{equation}
   In this line, the first $C_2$ is preceded by $\sigma - 1$ zeros and followed by $S - \sigma$ zeros.

   The estimated solution $\widehat X$ of the problem is the least-squares solution of this linear system.

\section{A performance assessment using DORIS data on GRAVES}
\label{sec:doris}
  \subsection{Method of assessment}
    In order to build a TEC model at the emission station and the reception station, we use DORIS-equipped satellites. The DORIS system~\citep{auriol_doris_2010} allows us to obtain (a posteriori) the orbit of some satellites with centimetric precision~\citep{jayles_ten_2004}. This in turn allows us to simulate the Doppler measurements that GRAVES should have produced, in the absence of any perturbation, according to the DORIS orbit reconstruction.
    So, on the one hand, we have the radial velocity data points that GRAVES should have produced on some satellites in the absence of perturbation.
    On the other hand, we have the raw radial velocity data points that GRAVES actually produced on those satellites.
    We call $\Delta v$ the difference between those radial velocities. Most of it is expected to be a consequence of the influence of the ionosphere.

    We will call this $\Delta v$ the $\Delta v$ without ionospheric correction. We will compare it to $\Delta v$ with ionospheric correction, that is, with the correction calculated in section~\ref{sec:gnsstotec}. Since the correction is applied to satellite in low Earth orbit, but the correction is calculated on GPS satellites, we apply a factor 0.9 to the correction, because about 10\% of the ionosphere is included between the altitdue of 1\,000 km (low Earth orbit) to 25\,000 km (orbit of GPS satellites) \citep{nava_new_2008}.

    In this paper, all $\Delta v$ are expressed in units of $\kappa$, which is homogeneous to m.s\textsuperscript{-1}, and which is the typical root mean square of the noise on radial velocity measurements for GRAVES (excluding ionospheric effects). \rev{We chose to express $\Delta v$ in units of $\kappa$ because the detailed performance of GRAVES is not public at the moment.}

  \subsection{The data}
    We consider the period from June 12\textsuperscript{th}, 2020, to July 6\textsuperscript{th}, 2020. We chose this period in particular because we had good data availability from the GPS ground stations over it. Some of the DORIS-equipped satellites on which we test our model are listed in table~\ref{tab:targets} along with their apsis. Over the whole period, we have a total number of 295 passes by DORIS satellites.

    \begin{table}[!htbp]
      \begin{tabular}{l l r r}
        NORAD number & name & perigee & apogee\\
        \hline
        36508 & Cryosat 2   & 723 km    & 730 km   \\
        37781 & Haiyang 2A  & 975 km    & 976 km   \\
        39086 & Saral       & 791 km    & 793 km   \\
        41335 & Sentinel-3A & 809 km    & 811 km
      \end{tabular}
      \caption{Target satellites, along with their NORAD identification number, altitude at perigee, and altitude at apogee.}
      \label{tab:targets}
    \end{table}

    The data on the GPS satellites are downloaded from the IGN website (\url{http://rgp.ign.fr/DONNEES/diffusion/}), using a 2 second sampling period, and excluding GPS satellites when their elevation is less than $10^{\circ}$. The stations whose data we used are DJON, in Dijon, France, and APT1 near Apt, France. Those stations are close to the emission segment (DJON) and the reception segment (APT1) of GRAVES.

\begin{figure}[!htbp]
      \includegraphics[width=\linewidth]{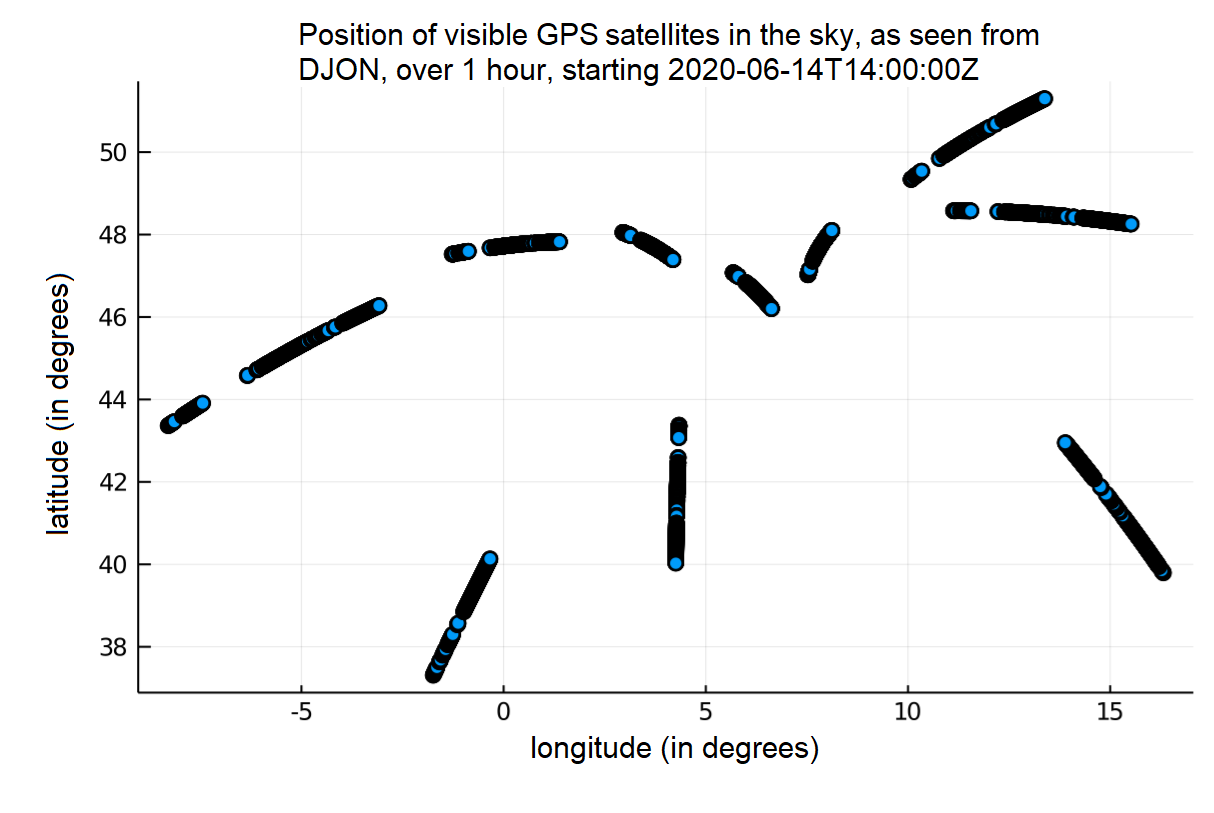}
      \caption{Representation of the position of GPS satellites in the sky seen from the DJON station with an elevation greater than $10^{\circ}$ on June 14\textsuperscript{th}, 2020, between 14:00 UTC and 15:00 UTC. It is quite representative of the general sky coverage of GPS as seen from a station in mainland France.}
      \label{fig:couverture_ciel}
    \end{figure}

    In order to get a feeling of the sky coverage of the GPS satellites as seen from a ground station, we present on figure~\ref{fig:couverture_ciel} the typical sky coverage by GPS satellites as seen from a ground station over the course of an hour.

  \subsection{Results over the whole period}
\begin{figure}[!htbp]
      \includegraphics[width=\linewidth]{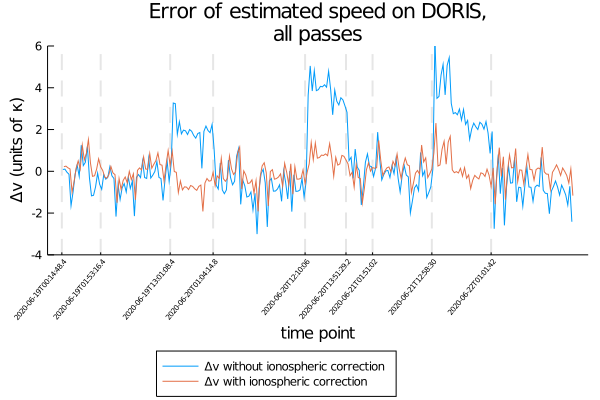}
      \caption{$\Delta v$ on DORIS for all passes (restricted to 250 data points)}
      \label{fig:all_passes}
    \end{figure}

    Figure~\ref{fig:all_passes} displays 250 data points of $\Delta v$ with or without correction. The correction visibly leads to a lower error, thus a better estimation of the velocity. \rev{On this figure, nine passes are displayed. The data is displayed continuously in order to facilitate visual interpretation, but there is obviously a time discontinuity between the end of each pass and the beginning of the next pass. The beginning of each pass is signaled by a vertical line, along with an indication of the time at the beginning of the pass.}

    Quantitatively, the root mean square averages of the velocity error $\Delta v$ with and without ionospheric correction are very different. Without ionospheric correction, the error is 2.28$\kappa$, while it is only 0.95$\kappa$ with ionospheric correction. The correction allows us to compensate 58\% of the error induced by the ionosphere.

    \rev{On this graph, we notice that correction seems to be much more necessary during daytime than during nighttime.} Since the main driver of the activity of the ionosphere is the Sun, we analysed in more details passes during the night and passes during the day.

  \subsection{Results on nighttime only}
\begin{figure}[!htbp]
      \includegraphics[width=\linewidth]{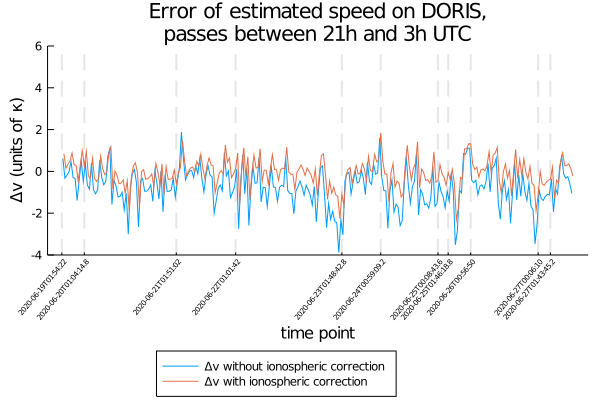}
      \caption{$\Delta v$ on DORIS for night passes (restricted to 250 data points)}
      \label{fig:night_passes}
    \end{figure}

    On June 20\textsuperscript{th}, 2020, sunrise occurred at 5:44 (local time) in Dijon, and sunset occurred at 21:38. Since the longitude of Apt is close to that of Dijon and Apt is South of Dijon, sunrise occurs in Apt after Dijon, and sunset occurs in Apt before Dijon around the summer solstice.
    For the purpose of this paper, we define “nigthtime” as the time between 21:00 and 3:00 UTC, which correspond to 23:00 to 5:00 local time. We chose those times in order to avoid twilight time as much as possible, while retaining enough data points to be able to compute statistical quantities. In this case, we have several thousand data points for nighttime.
    Figure~\ref{fig:night_passes} displays 250 data points of $\Delta v$ with or without correction. Although the correction leads to a lower error, the correction is less beneficial than when all passes are considered.

    \rev{Just after the end of twilight, the ionosphere had not yet had time to relax completely. Those are the occurrences when nighttime correction is particularly useful.}

    Quantitatively, the root mean square averages of the velocity error with and without ionospheric correction are quite close. Without ionospheric correction, the error is 1.53$\kappa$, and it is 1.09$\kappa$ with ionospheric correction. The correction allows us to compensate 29\% of the error induced by the ionosphere. It is not negligible, but it is less than half the correction for all passes.

  \subsection{Results on daytime only}
\begin{figure}[!htbp]
      \includegraphics[width=\linewidth]{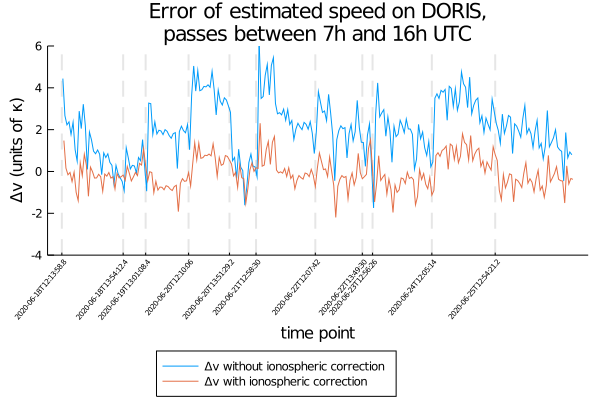}
      \caption{$\Delta v$ on DORIS for passes during daytime (restricted to 250 data points)}
      \label{fig:day_passes}
    \end{figure}

    For the purpose of this paper, we define “daytime” as the time between 7:00 and 16:00 UTC, which correspond to 9:00 to 18:00 local time. We chose those times in order for the sun to be quite quite high in the sky. In this case, we have several thousand data points for daytime.
    Figure~\ref{fig:day_passes} displays 250 data points of $\Delta v$ with or without correction. It is clear that the correction during the day is extremely useful.

    Quantitatively, the root mean square averages of the velocity error with and without ionospheric correction are very different. Without ionospheric correction, the error is 2.60$\kappa$, and it is 0.83$\kappa$ with ionospheric correction. The correction allows us to compensate 68\%, more than 1.7$\kappa$ RMS of the error induced by the ionosphere.

\rev{
\section{Discussion and perspectives}
  \subsection{Discussion of the results}
    The previous quantitative results are brought together for comparison in table~\ref{tab:rms_synthese}.

    \begin{table}[!htbp]
      \begin{tabular}{l r r}
          $\Delta v$ (RMS) & without correction & with correction \\
          \hline
           all passes & 2.28$\kappa$ & 0.95$\kappa$\\
            nighttime passes & 1.53$\kappa$ & 1.09$\kappa$ \\
            daytime passes & 2.60$\kappa$ & 0.83$\kappa$
      \end{tabular}
      \caption{Root mean square averages of the velocity error with and without ionospheric correction for all passes, for passes during nighttime, and for passes during daytime.}
      \label{tab:rms_synthese}
    \end{table}
    First, we recall that $\kappa$ is the typical root mean square of the noise on radial velocity measurements for GRAVES excluding ionospheric effects, while $\Delta v$ is the root mean square of the noise on radial velocity measurements due to ionospheric effects.
    Since $\Delta v > \kappa$, ionospheric effects are clearly an important contributor to the overall error, thus it is indeed important to correct them, as stated in the introduction.

    We can see that ionospheric effects are more important during daytime than during nighttime. This is not surprising, as the Sun is largely the main driver of ionospheric activity.

    Also, while our proposed correction works during nighttime, it is during daytime that it is both most efficient and most necessary.

  \subsection{Limitations and perspectives}
    The method presented here can be refined in various ways. Instead of using the simple standard single layer model for equation~\ref{eq:modeledirect}, we could use a modified single layer model mapping as in \cite{feltens_tropospheric_2018}. We could also constrain the reconstruction of the parameters like in \cite{yasyukevich_gnss-based_2020}.

    We could test our reconstructions with a more general formula to take into account the altitude of the satellite, as well as the climatological state of the ionosphere, which is more contracted during solar minimum and expanded during solar maximum.

    Finally, we did not investigate the influence of solar activity (which is responsible for extreme ultraviolet photons production) nor geomagnetic activity on the ionosphere, while these are well-known factors. We expect the correction of ionospheric effects to be more important during episodes which induce high ionospheric activity, and we will have to study this during the next solar maximum, in a few years.

}

\section{Conclusion}
We developed a simple method to address the need for taking into account ionospheric effects in the use of a space surveillance radar. This method relies only on data produced by dual-frequency GNSS \rev{receivers} at the emission and reception sites. We tested this method on GRAVES data using DORIS a posteriori orbits as ground truth. This method allows a notable reduction of the error due to the ionospheric effect. In particular, we reach a reduction in the radial velocity estimate from 2.60$\kappa$ to 0.83$\kappa$ (root mean square average) on daytime measurements.

 So, while the method that we propose is very simple to implement, it has proven to be a good correction for ionospheric effects in practice.

\section*{Acknowledgements}
We thank the \textit{Direction g\'en\'erale de l'armement} of the French \textit{Minist\`ere des Arm\'ees} for having funded the GRAVES CGC2 TC2 study, with financial support from the European Space Surveillance and Tracking Programme (EUSST). We thank in particular our correspondents at the \textit{Direction g\'en\'erale de l'armement} for their support and for accepting our request to publish some of our results. We thank Florent Muller for his support and valuable advice. We thank Emmanuelle Laneel for her assistance with the project management of this study. \rev{We thank the reviewers for their very constructive criticism and valuable comments that greatly helped to improve
this paper.}

\bibliographystyle{model5-names}
\biboptions{authoryear}
\bibliography{biblio}

\end{document}